# Uncovering Cellular Resolution in scRNAseq via Unbiased Cell and Gene Network Analysis

**Olga Ianzetta, Department of Experimental Medicine and Internal Medicine, University of Genova, Italy**

**Luisa Cutillo, School of Mathematics, University of Leeds, UK**

**Bailey Andrew, Division of Cancer Sciences, University of Manchester, UK**

**Claudia Angelini, Istituto per le Applicazioni del Calcolo, IAC-CNR, Italy**

## Abstract

Conventional annotation of single-cell RNA-sequencing (scRNA-seq) data relies heavily on manual, marker-based thresholding, an approach that can obscure subtle transcriptomic gradients and collapse functionally distinct cell states into broad, heterogeneous populations. Here we apply the Gaussian multi-Graphical Model (GmGM) framework, which jointly infers cell-cell and gene-gene dependency structure from a single scRNA-seq data matrix, to a 10x Genomics PBMC dataset. Ten independent GMGM-Leiden clustering runs were integrated into a robust consensus partition using a soft cluster ensemble approach and benchmarked against reference cell-type annotations. This strategy yielded stable cluster partitions that resolve biologically meaningful sub-populations not distinguished by the reference annotation. In parallel, for each cluster, gene co-expression modules were extracted from the fitted model via consensus Leiden clustering across resolutions, evaluated using standard network metrics, and validated functionally with the Network Enrichment Analysis Test (NEAT), which confirmed non-random enrichment signal. A module-scoring procedure linked network topology to per-cell, per-cluster expression signatures, and a novel extension of GmGM, recovering a shared cell-cell network together with population-specific gene networks in a single model run, was demonstrated in a case study on the CD4+ T-cell population. These results indicate that GmGM provides a unified, reproducible framework for joint cell clustering and gene-network inference, capable of revealing cellular structure beyond that captured by conventional pipelines.

# 1. Introduction

## 1.1 Background and Rationale

Conventional scRNA-seq annotation relies heavily on manual, marker-based thresholding, an approach prone to obscuring subtle transcriptomic gradients and to collapsing functionally distinct cell states into broad, heterogeneous macro-populations. This motivates the adoption of an unsupervised, network-driven approach: the Gaussian multi-Graphical Model [Andrew et al. 2026], benchmarked directly against standard reference cell-type annotation, with the specific aim of recovering hidden cellular resolution that conventional pipelines may fail to capture.

## 1.2 Objectives

This report describes the analytical workflow developed for the application of the GmGM framework to scRNA-seq data derived from human peripheral blood mononuclear cells (PBMCs). The study pursues two complementary objectives: (i) to benchmark the ability of GmGM to recover cell-type clusters concordant with an established reference annotation; and (ii) to exploit the gene co-expression networks inferred by the model to identify functionally coherent gene modules, characterize them topologically and biologically, and extend the framework to population-specific gene networks.

# 2. Materials and Methods

## 2.1 Dataset

The dataset comprised $N = 10{,}985$ high-quality single-cell RNA-seq profiles of human peripheral blood mononuclear cells (PBMCs) from a healthy donor, generated using Chromium Next GEM Single Cell 3' v3.1 single-indexed chemistry (10x Genomics) and sequenced on an Illumina NovaSeq 6000 platform. This dataset was processed and utilized as a reference benchmark for scRNA-seq computational evaluation as described by Kharchenko (2021).

## 2.2 The GmGM Model

GmGM jointly estimates gene and cell dependencies from a single data matrix

$$\mathrm{vec}(D) \sim N\left(0, \left(\Psi_{\mathrm{gene}} \oplus \Psi_{\mathrm{cell}}\right)^{-1}\right)$$

Where $\Psi$ denotes the precision matrix.

Gram matrices, computed directly from the raw data as sufficient statistics, capture pairwise correlation structure across genes and across cells:

$$S_{\mathrm{gene}} = D^T D, S_{\mathrm{cell}} = DD^T$$

GmGM exploits the fact that the eigenvectors V of the maximum likelihood estimate of the precision matrices coincide with those of the Gram matrices (Theorem 1, Andrew et al. 2026). Optimization is therefore restricted to the eigenvalues $\Lambda$ alone, yielding a large speed-up over prior multi-axis Gaussian graphical model formulations.

$$S_l = V_l \cdot \mathrm{diag}(e_l) \cdot V_l^T \Rightarrow V_l = \mathrm{eig}(\Psi_l), \Psi_l = V_l \Lambda_l {V_l}^T$$

where the eigenvalue $\Lambda$ is are solved iteratively.

The resulting conditional dependency graph is defined by the non-zero entries of $\Psi$ after sparsification:

$$\Psi_{ij} = 0 \Leftrightarrow i \perp j \mid \text{rest of data}$$

These non-zero entries isolate direct partial correlations, explicitly defining the edges for both the cell-cell and gene-gene networks utilized in downstream analyses.

As GmGM only depends on the sufficient statistics, it can be expressed as a function $\mathrm{GmGM}(S_{\mathrm{cells}}, S_{\mathrm{genes}})$. GmGM can also work for datasets with 'shared axes', such as paired scRNA-seq and scATAC-seq datasets, in which both modalities describe the same set of cells (the 'shared axis'). scATAC-seq features are called 'peaks'; applying GmGM to this type of dataset be expressed as $\mathrm{GmGM}(S_{\mathrm{cells}}, S_{\mathrm{genes}}, S_{\mathrm{peaks}})$.

## 2.3 Analytical Workflow Overview

The pipeline comprises the following stages: (i) input ingestion and quality control of the scRNA-seq dataset; (ii) GmGM inference, in which precision-matrix estimation jointly yields a gene-gene network (direct partial correlations) and cell-cell precision matrices, composing a multi-axis graphical model; (iii) cell clustering, via 10 parallel Leiden runs (seeds 1-10) at different resolutions; (iv) consensus integration across runs; (v) comparative evaluation against reference cell-type annotations via a generalized, column-normalized confusion matrix; and (vi) validation and sub-population discovery, through refinement of clusters informed by the identified gene modules.

## 2.4 Cell Clustering and Consensus Integration

Cell clustering was performed by applying the Leiden algorithm to the cell-cell adjacency matrix derived from the fitted model, with the procedure repeated 10 times under distinct random seeds, allowing the robustness of the clustering solution to be assessed with respect to algorithmic stochasticity. Each run was benchmarked against the reference cell-type annotations using the Adjusted Rand Index (ARI) and Normalized Mutual Information (NMI). To synthesize a single, robust partition across the 10 independent runs, a cluster ensemble consensus strategy was implemented using the clue R package (Hornik, 2005). The individual clustering outputs were integrated using a soft consensus optimization (method = \"SE\") to resolve run-to-run variability and derive stable cell assignments. Concordance between the consensus partition and reference cell-type annotation was further examined via a generalized, column-normalized confusion matrix relating the resulting GmGM clusters to reference cell types.

## 2.5 Population-Specific Gene Networks: an Extension of GmGM

GmGM jointly learns the cell precision matrix ($\Psi_{\mathrm{cells}}$) and the gene precision matrix ($\Psi_{\mathrm{gene}}$) from a single data matrix $D$. However, in the presence of multiple cell types (e.g., $\mathrm{Cl}_1$, $\mathrm{Cl}_2$, $\mathrm{Cl}_3, \cdots$), distinct gene networks are often required for each population rather than a single global gene network, which in principle would require recomputing the gene network separately for every population. Not only does this require running GmGM several times, but such an approach is unable to take into account knowledge gained from similar cell types – CD4 T cells and CD8 T cells are related cell types, but the optimization problem for a CD8-specific gene network does not depend at all on the information contained in CD4 T cells. Being able to jointly learn all networks would improve performance, especially when working with rare cell types, in which case information from similar cell types may be necessary to reconstruct the network.

To address this, we can conceptualize our data as coming from a 'virtual dataset' consisting of multiple matrices, one for each cell type, that all share the cell axis but have a separate gene axis. While it is not possible to physically construct such a dataset, it *is* possible to construct the Gram matrices it would have if it were constructable – these are given below.

$$D = \left[D_{\mathrm{Cl}_1} D_{\mathrm{Cl}_2} D_{\mathrm{Cl}_3} \cdots\right], S_{\mathrm{cell}} = DD^T, S_{\mathrm{gene,Cl}_k} = D_{\mathrm{Cl}_k}^T D_{\mathrm{Cl}_k}$$

As GmGM only depends on the Gram matrices, it can still be applied: $GmGM(S_{cell}; S_{gene,\,Cl_1}; S_{gene,\,Cl_2}; \cdots)$. The output of a single model run therefore consists of a global cell network ($\Psi_{\mathrm{cell}}$) together with a distinct gene network per population ($\Psi_{\mathrm{gene,\,Cl}_k}$), eliminating the need to run the model separately for each subgroup.

The principal limitation of this approach is that cell types must be known a priori: if clusters are instead discovered using GmGM itself, the strategy must be rerun post hoc using the discovered clusters as input.

### 2.6 Gene Network Extraction and Module Detection

The cluster-specific gene co-expression network was extracted from the fitted GmGM model. Leiden communities were computed on the gene network across more resolution values, with 20 independent runs per resolution obtained by varying the random seed at each iteration. Consensus clustering was then used to obtain a stable partition at each resolution. For each consensus partition, the number of clusters, minimum and maximum module size, average degree, clustering coefficient, and network modularity were computed to guide selection of the optimal resolution, after which modules were ranked within each cell type.

### 2.7 Functional Enrichment Analysis (NEAT)

The biological relevance of the identified gene modules was assessed using the Network Enrichment Analysis Test (NEAT; Signorelli & Cutillo, 2022), which tests whether genes within a module are more strongly connected to a reference gene list than expected by chance, explicitly accounting for the topology of the background network, a substantially more informative approach than simple list-overlap analysis.

The test requires three principal inputs: (i) a reference gene list, i.e. a curated, well-characterized set of genes of interest (e.g., a pathway or a disease signature); (ii) the module assignment, i.e. the gene modules identified via clustering, one set per resolution or consensus partition; and (iii) the background network, i.e. the full gene network used to define the null model of expected connectivity. The test output is an FDR-adjusted enrichment p-value per module, allowing identification of the module in which the reference gene list is most strongly enriched.

### 2.8 Gene-Module Scoring by Cell Cluster

To link the gene modules identified on the network to observed cellular heterogeneity, the following procedure was implemented: (i) identification of cell clusters and gene modules via Leiden clustering on the network; (ii) computation, per cell, of the percentage of raw expression attributable to each gene module; (iii) construction of a new object in which each row still represents a cell, but columns now represent individual gene modules; and (iv) identify gene modules significantly associated with each cell cluster. This strategy reduces the topological information encoded in the gene network (i.e., the modules) to a per-cell, per-population interpretable quantity, conceptually analogous to a module score.

## 3. Results

### 3.1 Clustering Performance Across Runs and Consensus Partition

Results indicate strong cross-run consistency: ARI ranges from 0.7394 to 0.8237, and NMI ranges from 0.8237 to 0.8676, reflecting a high and stable level of agreement between independent Leiden runs and the reference cell-type annotations (Table 1).

| Run | ARI | NMI |
|---|---|---|
| 0 | 0.7404 | 0.8244 |
| 1 | 0.7966 | 0.8557 |
| 2 | 0.8144 | 0.8484 |
| 3 | 0.8186 | 0.8472 |
| 4 | 0.7997 | 0.8570 |
| 5 | 0.8237 | 0.8676 |

| Run | ARI | NMI |
|---|---|---|
| 6 | 0.7987 | 0.8549 |
| 7 | 0.7948 | 0.8370 |
| 8 | 0.7394 | 0.8237 |
| 9 | 0.8224 | 0.8665 |

*Table 1. ARI and NMI indices computed for each of the 10 Leiden algorithm runs relative to the reference cell-type annotation.*

The consensus partition was computed on the 10 individual runs, setting the clustering resolution parameter to 0.8. The resulting consensus partition exhibits high inter-run agreement, corroborating the stability and reproducibility of GmGM-derived cellular communities. Benchmarking this consensus partition against reference cell-type annotation confirms overall clustering quality (Table 2), with performance metrics consistent with the mean across individual runs.

| Comparison | ARI | NMI |
|---|---|---|
| Consensus | 0.7981 | 0.8557 |

*Table 2. ARI and NMI indices for the consensus-clustering partition relative to reference cell-type annotation.*

### 3.2 Cell Population Identification

The correspondence between the GmGM consensus partition and reference cell-type annotation was further examined via a generalized, column-normalized confusion matrix, relating the 14 GmGM clusters to reference cell-type annotations (Fig.1) The matrix reveals a strong, largely one-to-one correspondence for the majority of clusters, with residual overlap concentrated among closely related populations (e.g., monocyte or T-cell subsets), a pattern more consistent with transitional cell states than with genuine misclassification (Fig.2).

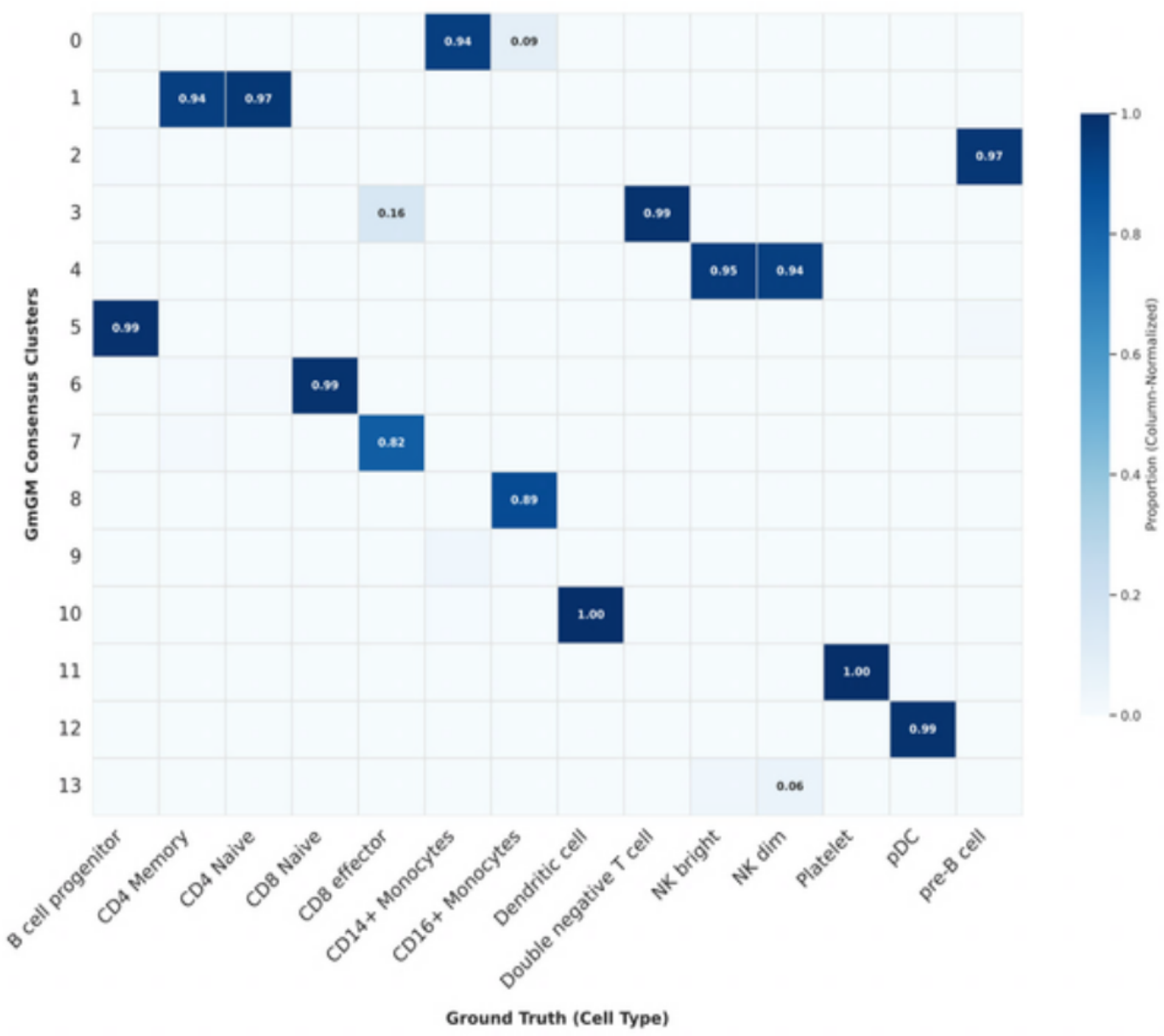

*Fig.1 Column-normalized confusion matrix, relating the 14 GmGM clusters to reference cell-type annotation*

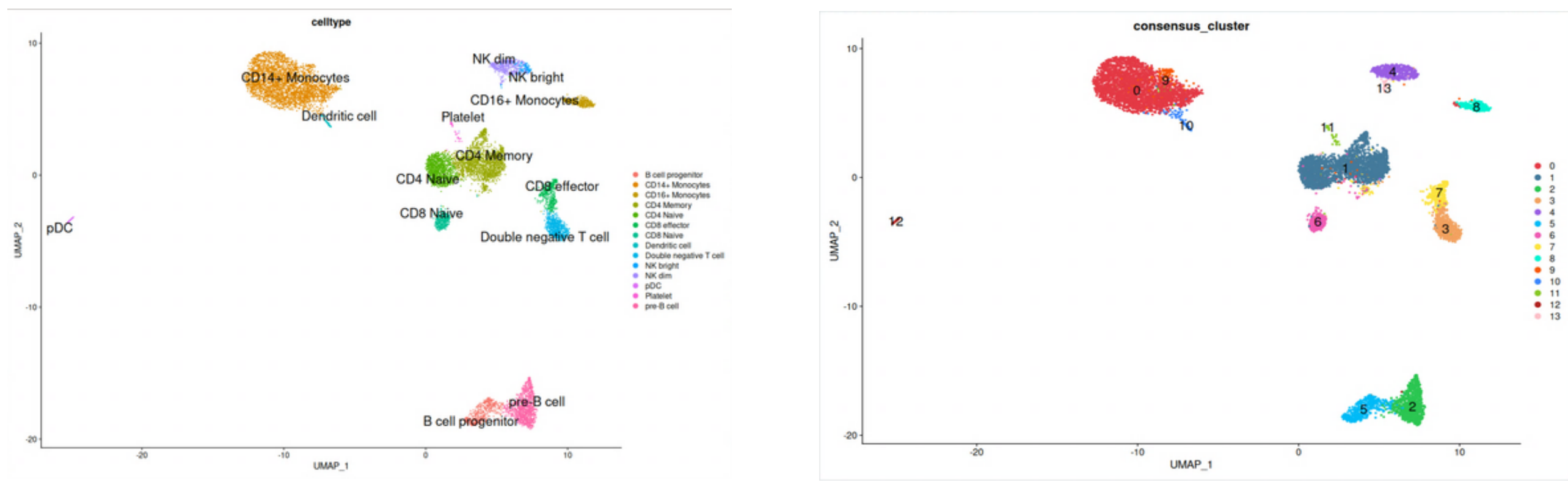

*Figure 2. Side-by-side comparison of reference cell-type annotations and GmGM-derived consensus clusters.*

Marker-gene analysis performed on the consensus partition confirms high biological consistency across clusters, and simultaneously highlights that conventional cell-type annotation annotation can itself be limited by insufficient resolution relative to what the model is capable of resolving. Table 3 reports the cell population assigned to each of the 14 consensus clusters, based on the most differentially expressed marker genes.

| Cluster | Identified Cell Population |
|---|---|
| 0 | Classical Monocytes |
| 1 | Naive CD4+ T Cells |
| 2 | Naive B Cells |
| 3 | Effector Memory CD8+ T Cells |
| 4 | Natural Killer (NK) Cells |
| 5 | Plasma Cells / Plasmablasts |
| 6 | Naive CD8+ T Cells |
| 7 | Terminal Cytotoxic CD8+ T Cells |
| 8 | Non-Classical Monocytes |
| 9 | IFN-Induced Myeloid Cells |
| 10 | Conventional Dendritic Cells (cDC2) |
| 11 | Erythroid Contamination |
| 12 | Plasmacytoid Dendritic Cells (pDC) |
| 13 | Proliferating Cells |

*Table 3. Cell population assigned to each consensus cluster, based on marker-gene analysis.*

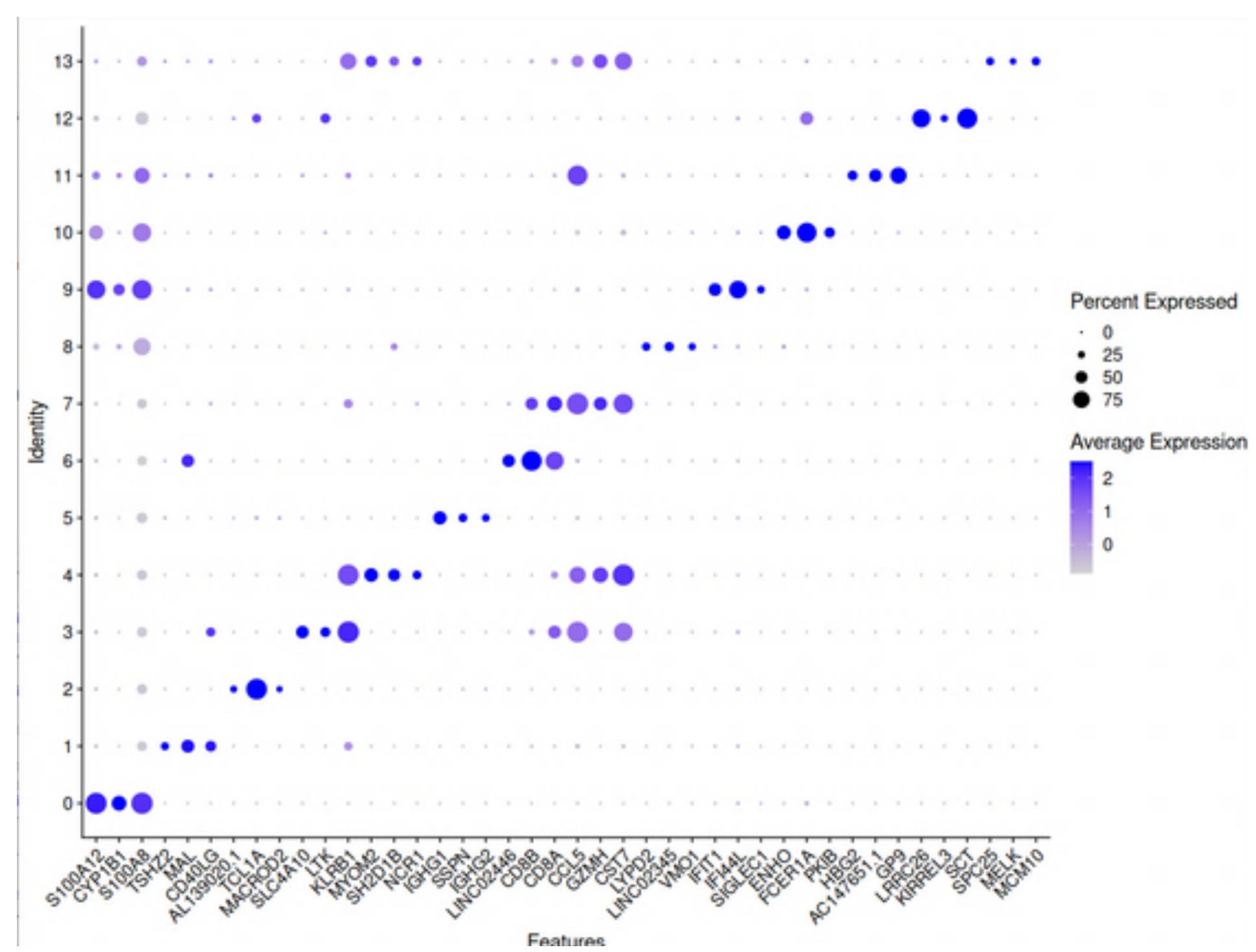

*Figure 2. Dot plot of marker-gene expression by cluster: dot size = percentage of cells expressing the gene; color = average expression level.*

Comparison of GmGM clusters with reference cell-type annotations yields three principal observations: (i) discrepancies between the two annotation schemes are not necessarily classification errors, and frequently reflect genuine underlying biological complexity; (ii) clusters exhibiting partial overlap tend to capture transitional cell states and shared marker expression between closely related populations; and (iii) GmGM successfully resolves novel sub-populations within cell types that reference cell-type annotation treats as homogeneous; for instance, the separation of classical, non-classical, and IFN-induced monocyte subsets, or the identification of a discrete proliferating-cell component.

### 3.3 Selection of the Optimal Gene-Module Resolution

As a practical application of the workflow, the analysis focused on the CD4+ cell population (corresponding to Cluster 1), with the aim of stabilizing this population's gene modules across resolutions. The procedure comprised six stages: (1) selection of gene network for Cluster 1, corresponding to the CD4+ population; (2) identify the modules across a range of resolutions from 0.4 to 0.9; (3) consensus clustering, combining the resolution-wise partitions into a single consensus clustering using the R package clue; (4) modularity assessment, computing modularity and network metrics for each resolution/consensus result to evaluate partition quality; (5) selection of the resolution (or consensus partition) offering the best modularity and the most stable modules and (6) functional enrichment of gene modules.

| Resolution | N. Modules | Min Size | Max Size | Avg Degree | Clustering Coeff. | Modularity |
|---|---|---|---|---|---|---|
| 0.4 | 11 | 1 | 1225 | 3.296 | 0.1396 | 0.6664 |
| 0.5 | 15 | 4 | 1195 | 3.277 | 0.1468 | 0.6991 |
| 0.6 | 18 | 7 | 1204 | 3.224 | 0.1553 | 0.7148 |
| 0.7 | 22 | 7 | 1202 | 3.058 | 0.1630 | 0.7121 |
| 0.8 | 22 | 35 | 1202 | 3.097 | 0.1529 | 0.7223 |
| 0.9 | 26 | 22 | 1206 | 3.094 | 0.1624 | 0.7201 |

*Table 5. Network metrics computed for consensus partitions across resolutions, restricted to the CD4+ population gene network.*

As in the global-network analysis, the number of modules increases with resolution (from 11 to 26), with modularity values in the range 0.66-0.72 and a maximum at resolution 0.8 (0.7223). Relative to the global gene-network analysis (Table 4), modules identified within the CD4+ population exhibit a markedly lower average degree (approximately 3.0-3.3 versus 5.8-7.5) and a lower clustering coefficient, consistent with a more circumscribed network specific to this single cell population.

### 3.4 Functional Enrichment of Gene Modules

Applying NEAT across the identified modules, out of 26 total comparisons, 2 enrichments were significant at the 1% level and 8 at the 5% level. The Kolmogorov-Smirnov test for uniformity of p-values returned a value of 0, indicating that the observed p-value distribution departs significantly from the uniform distribution expected under the null hypothesis, supporting the presence of a genuine, non-random enrichment signal.

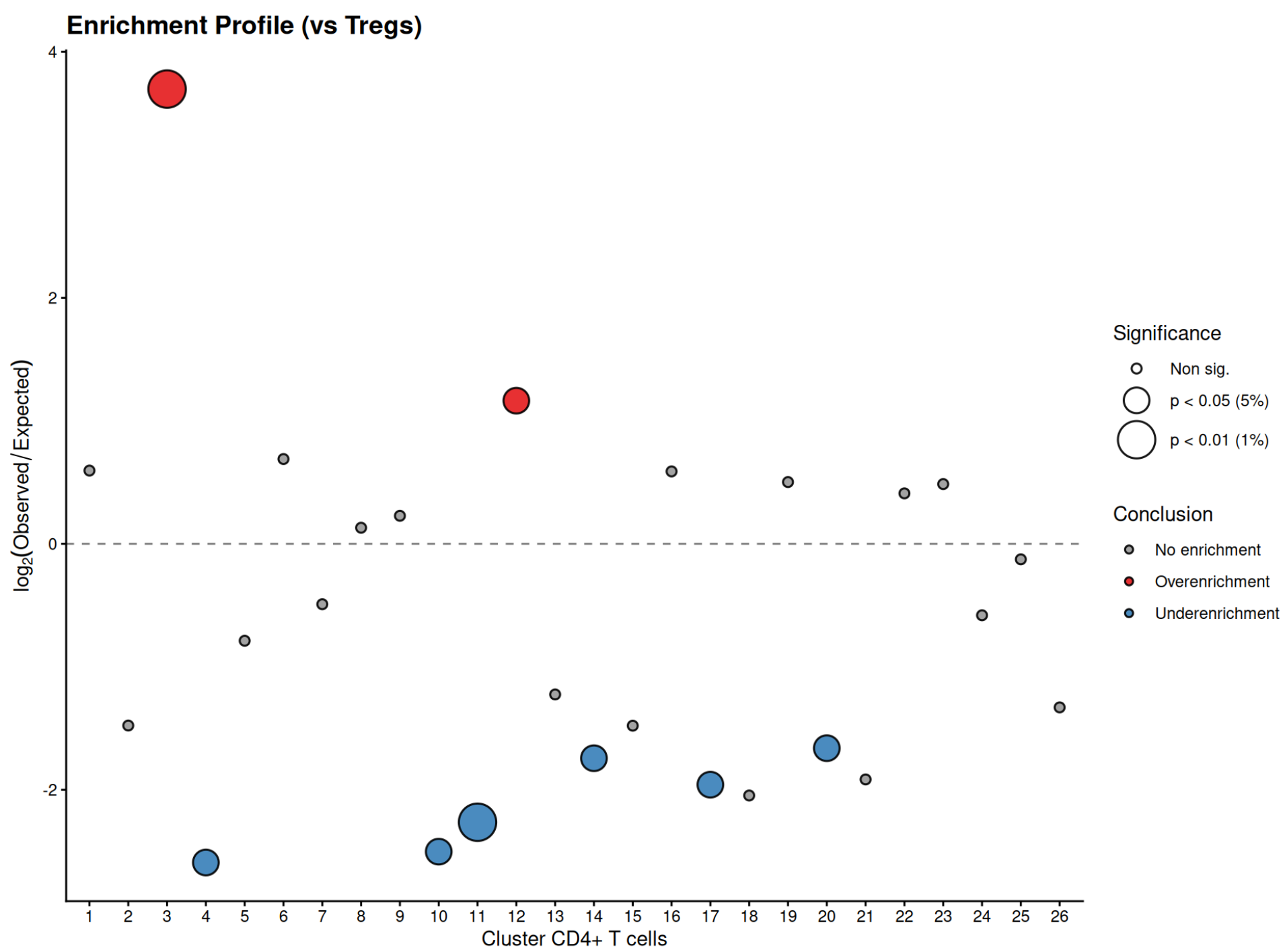

*Figure 3. Spatial Enrichment Profile of CD4+ T Cell Subpopulations Relative to Tregs. Dot plot representing the spatial co-localization analysis between each CD4+ T cell cluster (X-axis) and Regulatory T cells (Tregs). The Y-axis depicts the log2-transformed ratio of observed to expected pairwise cell interactions (log2(Observed / Expected)). The horizontal dashed line (Y = 0) marks spatial randomness. Circle colours indicate the enrichment conclusion based on adjusted p-values: red for statistically significant overenrichment (spatial attraction), blue for significant underenrichment (spatial avoidance), and grey for no significant spatial deviation. Circle sizes denote the significance threshold: small (non-significant, p >= 0.05), medium (p < 0.05), and large (p < 0.01).*

## 4. Discussion

Taken together, these results indicate that GmGM reconstructs, from scRNA-seq PBMC data, cell clusters that are both robust and biologically consistent, providing a unified and reproducible framework for single-cell analysis. Beyond validating clustering performance against reference cell-type annotation, GmGM identifies additional cellular structure that is otherwise obscured by conventional annotation, thereby generating novel biological hypotheses: most notably the separation of monocyte subsets and the identification of a discrete proliferating-cell component that the reference annotation does not resolve.

In parallel, the gene networks inferred by the model support the derivation of stable functional modules via consensus clustering, whose biological relevance is corroborated by NEAT enrichment testing and by module-scoring strategies linking the topological structure of the gene network to observed cellular heterogeneity.

The extension to population-specific gene networks, applied here as a case study to the CD4+ population, opens the possibility of characterizing gene regulation in specific cellular subtypes in a targeted manner, though this requires prior knowledge of the cell types of interest and remains, at present, an unpublished technique that awaits validation on additional datasets.

## 5. Conclusions and Future Directions

The workflow presented here demonstrates that the GmGM model is capable of reconstructing, from scRNA-seq PBMC data, cell clusters that are both robust and biologically consistent, while simultaneously supporting the derivation of stable, functionally validated gene co-expression modules. The following next steps have been identified for continuation of the project:

- Validation of the population-specific network strategy (currently unpublished) on additional datasets.
- Extension of enrichment analysis to a broader set of reference gene lists.
- Systematic optimization of clustering resolution as a function of cell type under analysis.